\documentclass[cameraready]{Interspeech}

\usepackage{array}
\usepackage{soul}
\sethlcolor{yellow}
\usepackage{hyperref}

\let\oldtabular\tabular
\renewcommand{\tabular}{\small\oldtabular}

\title{An Acoustic Landmark Database of the English Lexicon \\ via Articulatory Synthesis}

\author[affiliation={1,2}, correspondingauthor]{Mateo}{C\'{a}mara}
\author[affiliation={1}]{Jos\'{e} Luis}{Blanco}
\author[affiliation={3}]{Juan Ignacio}{Godino-Llorente}
\author[affiliation={2}]{Jeung-Yoon}{Choi}
\author[affiliation={2}]{Stefanie}{Shattuck-Hufnagel}

\address{
    $^1$ Signal Processing Applications Group, Information Processing \& Telecomm. Center, Universidad Polit\'{e}cnica de Madrid, Spain \\
    $^2$ Speech Communication Group, Research Laboratory of Electronics, Massachusetts Institute of Technology, USA \\
    $^3$ Bioengineering and Optoelectronics Lab., Universidad Polit\'{e}cnica de Madrid, Spain
}

\email{mateo.camara@upm.es, jl.blanco@upm.es, ignacio.godino@upm.es, jyechoi@mit.edu, sshuf@mit.edu}

\keywords{Acoustic landmarks, speech synthesis, articulatory synthesis, acoustic phonetics, Pink Trombone, computational phonetics.}

\begin{document}

\maketitle

\begin{abstract}
Acoustic landmark theory treats speech as organized around the acoustic consequences of articulatory gestures that shape the vocal tract and airflow. Progress is limited by the scarcity of large, unambiguously annotated landmark datasets. We invert the problem by generating speech from landmark patterns. Using the Pink Trombone physical vocal-tract synthesizer, we produce an English lexicon for two adult configurations (male, female). With direct control of gestures, we place landmark labels algorithmically at the exact times of their physical events (e.g., oral closures/releases). The corpus contains $>$200,000 synthesized words, rendered for both configurations with time-aligned annotations; intelligibility is measured with STOI. We leverage it for statistics across the lexicon from an articulatory-event view, reporting landmark frequencies and dominant cue patterns, and enabling quantitative studies plus training/benchmarking of automatic landmark detectors.
\end{abstract}

\section{Introduction}
\label{sec:intro}

The prevailing paradigm in Automatic Speech Recognition (ASR) processes the speech signal using uniform, frame-based analysis. However, insights from human speech perception suggest a more event-driven approach \cite{stevens_evidence_1981}. The Acoustic Landmark Theory, a foundational contribution to acoustic phonetics proposed by Kenneth Stevens, holds that speech is organized around discrete, perceptually salient events corresponding to critical articulatory gestures \cite{stevens_acoustic_2000, stevens_toward_2002}. These acoustic landmarks (such as the abrupt increase in spectral energy happening when the release of a stop consonant or the sonority peak of a vowel) serve as anchors for decoding the phonetic content of an utterance.

The event-based perspective suggested by the Landmark approach has substantial applications. Research has shown that speech frames containing landmarks are significantly more informative for ASR tasks than surrounding frames. Also, leveraging landmarks can improve accuracy, especially under challenging conditions \cite{he_improved_2018, he_selecting_2017}. Beyond ASR, landmark analysis has become a powerful tool in clinical phonetics. They provide objective metrics to assess speech disorders \cite{huang_investigation_2022, liu_first_2021}. For example, the SpeechMark toolkit demonstrated that patterns of landmark expression can quantify the impact of voice pathologies on intelligibility \cite{boyce_speechmark_2013}.

Despite its broad utility, progress in landmark-based research has been persistently hampered by a fundamental challenge: the scarcity of large-scale, accurately annotated corpora. Manual annotation is prohibitively labor-intensive and subject to inter-annotator variability \cite{juneja_probabilistic_2008}. Although recent efforts like the Auto-Landmark project have made a significant contribution by releasing a landmark-annotated version of the TIMIT corpus, those annotations entirely rely on their automatic annotator{. The accuracy of these automatic annotations has not yet been independently validated} 
\cite{zhang_auto-landmark_2024}. Landmark annotation, even when carefully addressed, {has to deal with the fact that coarticulatory and reduction processes are pervasive in continuous speech.} {That is, in connected speech, the acoustic realization of landmarks may be modified, weakened, or entirely omitted due to articulatory overlap between adjacent segments, where the gestures for neighboring sounds temporally overlap and interact. This means that the ``expected'' acoustic events predicted by the underlying phonological sequence may not always be clearly identifiable in the signal, further complicating the task of establishing reliable ground-truth annotations from natural speech data.}

This paper addresses this gap by proposing a different approach. Instead of analyzing real speech, we generate {a landmark-generated speech corpus} from first principles using a physically-based model. We employ the Pink Trombone\footnote{\url{https://dood.al/pinktrombone/}} \cite{thapen2017pink} (PT), a vocal tract synthesizer, to systematically generate the entire English lexicon.\footnote{\url{http://people.umass.edu/nconstan/CMU-IPA/}} Because we directly control the synthesizer's articulatory parameters (e.g., tongue position, lip closure, velic opening), we can algorithmically place landmark labels {with temporal precision}, at the exact moments of the underlying articulatory events. {In this generative framework, the landmarks emerge directly from the articulatory commands that produce the speech signal, rather than being inferred post-hoc from the acoustics.}


This {single-word-sized} dataset provides a sandbox environment, free from human labeling error. It is meant to validate the acoustic consequences of landmark theory and to serve as a ground-truth for training and benchmarking future automatic landmark detection systems.

The remainder of the paper is as follows: Section \ref{sec:background} defines a theoretical background, Section \ref{sec:meth} delineates the methodology for creating the database, Section \ref{sec:results} shows the landmark-based statistical results of the English lexicon, Section \ref{sec:discussion} discusses the results, and Section \ref{sec:conclusion} concludes the paper.

\section{Background}
\label{sec:background}

\subsection{The Physical Grounding of Landmark Theory}
Acoustic Landmark Theory {motivates} a model of the first steps in the process by which a listener decodes the continuously-varying speech signal into a sequence of discrete, meaningful linguistic units. To grasp the theory, it is essential to understand {the relationship between a number of different aspects of the speech production-perception processes} \cite{slifka_acoustic_2006}.

\begin{itemize}
    \item \textbf{Distinctive Features} are the abstract, binary properties that define phonemes in the mental lexicon (e.g., `[+nasal]', `[-continuant]'). They represent the intended linguistic {contrasts.}
    \item \textbf{Articulatory Gestures} are the physical actions of the vocal tract components (tongue, lips, velum) that {manifest} these features. For example, the feature `[-continuant]' is realized by a gesture of complete closure at some point in the oral tract.
    \item \textbf{Acoustic Events} are the direct, physical consequences of these gestures on the sound wave. A complete closure gesture yields in an abrupt drop in acoustic energy, while its release produces a sudden burst of broadband noise. The gesture of maximizing the oral cavity for a vowel results in a peak of sonorant energy in the signal. 
    \item \textbf{Acoustic Landmarks} are the labels applied to these highly salient acoustic events. A landmark is therefore not an imaginary construct. It is {a concrete aspect of the speech signal}, {with its location in time and its type determined by} a measurable physical reality in the signal.  {A landmark time stamp marks} a tangible event in the speech signal that indicates a critical moment in {the sequence of continuous articulatory movements} \cite{stevens_toward_2002}.
\end{itemize}

The connection is explicit and direct. {For example}, a Vowel landmark ($<$V$>$) is placed at the point of maximal energy of a vowel. A Stop Closure landmark ($<$Sc$>$) is placed at the moment of abrupt energy drop corresponding to oral closure. Similar definitions can be found {for all Landmark types}. {Thus, the landmark is a theoretical anchor} with its location determined by a measurable physical reality in the signal (and in the tract configuration).

\subsection{Classical Landmark Annotation}
Before the prevalence of end-to-end deep learning, the detection of acoustic landmarks was approached from a signal processing perspective rooted in expert phonetic knowledge \cite{liu_landmark_1996}. This classical methodology forms the basis for human annotation and validation, involving the identification of specific and observable patterns in the speech spectrogram that serve as reliable evidence for the underlying {articulatory events and intended features.} Our work relies on this connection for its validation: the landmarks we generate algorithmically correspond to events that are verifiable through these classical acoustic cues.

While the specific set and definitions of landmarks have evolved since the theory's inception, this work adopts {an updated approach that} categorizes them into eight primary types: four pairs of consonantal events, alongside vowel and glide landmarks {\cite{huilgol_2019}}. This classification refines earlier event-based notations \cite{boyce_speechmark_2013} (such as simple voicing onsets like $<$$+$g$>$/$<$$-$g$>$) into a more detailed, articulatorily-grounded framework \cite{he_acoustic_2018}. Table \ref{tab:landmark_cues} details these types, including their physical articulatory cause and the corresponding acoustic evidence that a human annotator or a rule-based algorithm would search for in a spectrogram.

\begin{table*}[htbp]
\caption{Landmark Types with their Articulatory Causes and Classical Acoustic Cues.}
\centering
\resizebox{\textwidth}{!}{%
{
\begin{tabular}{|c|m{0.40\textwidth}|m{0.6\textwidth}|}
\hline
\textbf{Landmark} & \textbf{Articulatory Event (Physical Cause)} & \textbf{Acoustic Cue (Evidence in Spectrogram)} \\
\hline
\textbf{V} & Maximal vocal tract opening for a vowel nucleus. & Peak in low-frequency sonorant energy. Strong and clear formant structure. \\
\hline
\textbf{G} & Narrowing of the vocal tract without creating turbulence. & Minimum in sonorant energy between two vocalic peaks, with smooth formant transitions. \\
\hline
\textbf{Sc} & Complete blockage of oral airflow by an articulator. & Abrupt cessation of energy, creating a ``stop gap'' or silence in the spectrogram. \\
\hline
\textbf{Sr} & Sudden release of built-up air pressure from a stop closure. & {A sharp spike} of broadband energy (burst transient) following a stop gap. \\
\hline
\textbf{Fc} & Formation of a narrow constriction causing turbulent airflow. & Onset of sustained, high-frequency aperiodic noise. \\
\hline
\textbf{Fr} & Widening of the constriction, ceasing the turbulent airflow. & Offset of the frication noise. \\
\hline
\textbf{Nc} & Oral closure combined with an open velopharyngeal port. & Abrupt damping of upper formants and onset of a low-frequency nasal murmur. \\
\hline
\textbf{Nr} & Release of the oral closure, redirecting airflow orally. & Abrupt transition from nasal murmur to a vowel-like formant structure. \\
\hline
\end{tabular}
}}
\label{tab:landmark_cues}
\end{table*}

This evidence-based approach is how the landmark annotations in our generated data can be validated. Our interactive visualization tool makes this link transparent by simultaneously displaying the spectrogram (the ``evidence'') and a real-time animation of the vocal tract model (the ``physical cause''). This provides an intuitive and direct confirmation of the foundational relationship between articulation and acoustics.

\subsection{The Pink Trombone: a Controllable Physical Model}
The central tenet of our work requires a synthesizer that allows us to control the \textit{origin} (articulation) and generate the \textit{manifestation} (acoustic signal). For this purpose, we selected the PT \cite{thapen2017pink}, an open-source physical model of a human vocal tract, powered by mathematical formulations of the underlying processes. It excels in providing an intuitive yet powerful interface for manipulating low-level speech production mechanisms, {which results in an initial landmark-based synthesis which reflects the acoustic patterns of the underlying Consonant-Vowel structure of an utterance}.

The synthesizer models the vocal tract as a waveguide and provides direct control over its primary articulatory components. A detailed technical description of the model, including its parameters and modifications for gender diversity, is provided in \cite{camara2025parameter}. The controllable elements include:
\begin{itemize}
    \item \textbf{The Glottal Source:} Controls the fundamental frequency ($f_0$) and degree of voicing, simulating the vibration of the vocal folds.
    \item \textbf{The Tongue Body:} A two-parameter system controls the position and diameter of the tongue, defining the primary shape of the oral cavity and the location of major constrictions.
    \item \textbf{Oral Constrictions:} Additional parameters allow for explicit constrictions at locations corresponding to the lips and the soft palate, crucial for producing stops and fricatives.
    \item \textbf{The Velopharyngeal Port (Nasal Cavity):} A binary control allows the port to be opened or closed, enabling the production of nasal sounds by coupling the nasal cavity to the vocal tract.
\end{itemize}

Although more anatomically informed and complex synthesizers exist, such as the Maeda model \cite{maeda1990compensatory} or VocalTractLab (VTL) \cite{birkholz2013modeling}, which offer higher physiological fidelity at the cost of increased computational complexity, the PT was deliberately chosen. As discussed in \cite{camara2025parameter}, its primary advantages for research in articulatory analysis and synthesis are its computational efficiency and simplicity --particularly after the modifications that were introduced to accelerate the synthesis process \cite{camara2025parameter}-- as well as the ease of interpreting its results, and its highly intuitive Graphical User Interface (GUI).

\section{Methodology}
\label{sec:meth}

\begin{figure}[htbp]
\centerline{\includegraphics[width=\columnwidth]{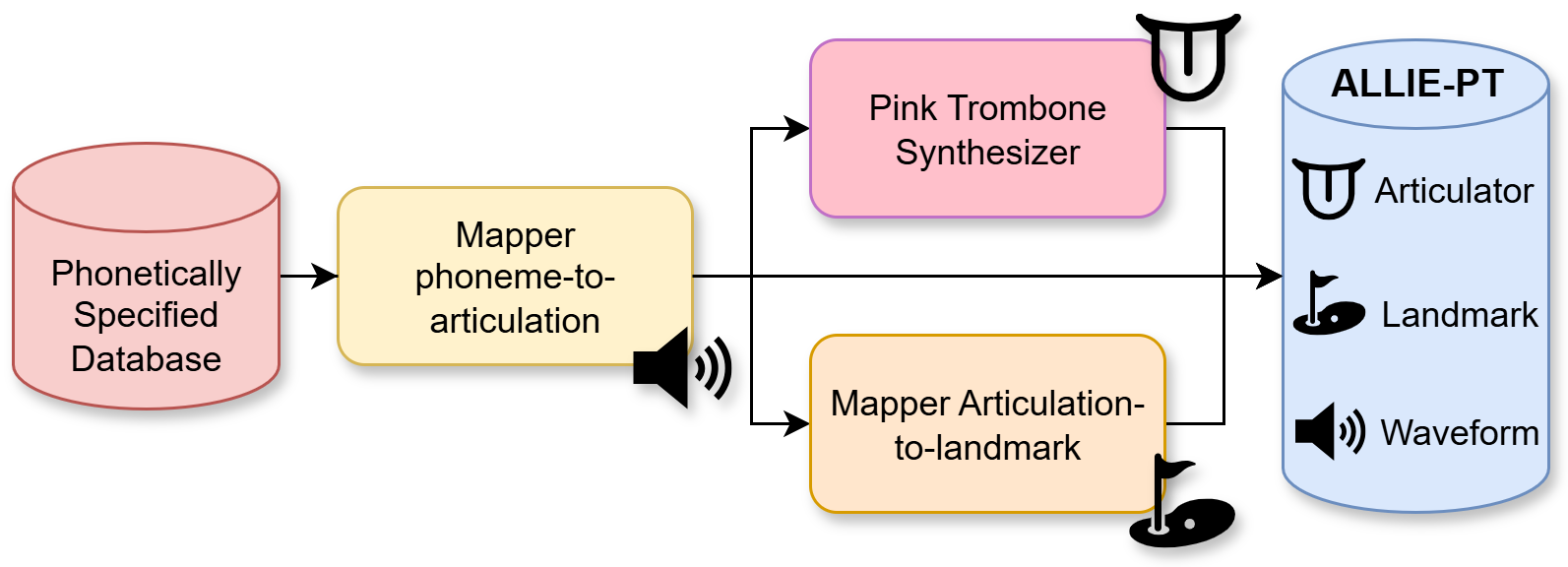}}
\caption{The data generation pipeline, from lexical entry to {the final landmark-specified waveform}.}
\label{fig:pipeline}
\end{figure}

Our process for generating the landmark-annotated database follows four main stages, as illustrated in Fig. \ref{fig:pipeline}.

\subsection{\texorpdfstring{{Linguistic Corpus and Phonological Transcription}}{Linguistic Corpus and Phonological Transcription}}
The foundation of our work is a comprehensive English lexicon derived from the Carnegie Mellon University Pronouncing Dictionary (CMUDict) \cite{cmudict}. The dictionary's original ARPABET transcriptions {of individual words} were converted to their {canonical representations} in the International Phonetic Alphabet (IPA) \cite{international1999handbook}, resulting in a master list of words and {their corresponding allophone sequences}. {(Although these sequences are described as phonemes in the database description, they are better characterized as phones, since they specify context-specific forms of the underlying context-invariant phonemes.)} This approach is grounded in prior work demonstrating that acoustic landmarks are robust and consistently {(although not invariably)} realized in American English \cite{shattuck-hufnagel_robustness_2007}. This suggest that bilingual corpora can be used on the assumption that landmark cues, which are tightly tied to the articulatory configuration that produce them, will not differ significantly in different languages. Preliminary results for Korean \cite{yun_2025_korean_labeling} and Italian \cite{di2021lexical} provide support for this assumption.

\subsection{Phoneme-to-Articulatory Mapping}
A crucial step was to define a mapping from each IPA phone to a static, canonical configuration of the PT's articulatory parameters. We mapped each English phoneme to targets in the synthesizer's control space (e.g., tongue-body height/advancement, tongue-tip position, lip rounding, jaw opening, velum, and glottal state), assembled manually from standard acoustic/articulatory descriptions and IPA sources \cite{stevens_acoustic_2000, maeda1990compensatory, birkholz2013modeling, international1999handbook}. Vowel targets follow the cardinal-vowel geometry. We then nudged parameters to place tokens within expected adult $F_1$–$F_2$ regions (±100 Hz). Diphthongs are implemented as linear trajectories between the two vowel targets with brief steady states at the endpoints (${\approx}$80--120 ms). For obstruents, we set the place and degree of constriction, voicing, and velopharyngeal state. Stops use a closure–burst template, fricatives use a narrow constriction to elicit turbulence, and nasals lower the velum at the canonical place. Liquids and glides use less extreme constrictions at the standard places (e.g., alveolar lateral for /l/, postalveolar rhotic for /\textturnr/). Targets correspond to canonical forms as if spoken alone, {not conditioned by phrase-level factors such as adjacent words and variable prosodic structure}. {In these word-sized sequences, we approximate a limited set of coarticulatory effects, such as changes in amplitude during movement toward closures and after releases,} via linear interpolation of articulatory parameters. Two authors cross-checked all entries and made small adjustments{, guided by listening and visual inspection of the spectrogram}. The final {phone-target} lookup and timing templates are released with the dataset card.

\subsection{Speech Synthesis and Coarticulation}
With the target configurations defined, word-level synthesis was automated. {For a given phone sequence} (e.g., {[k] [æ] [t]} for ``cat''), the system retrieves the corresponding articulatory target vectors. To model coarticulation and create a smooth utterance, we used linear interpolation between the parameter vectors of {adjacent phones}. This approach {has previously been validated} \cite{camara2023optimization}.

{Phone duration} was kept constant across all instances to create our controlled, aprosodic database. This deliberate choice removes prosodic variability, allowing the dataset to serve as a clean baseline for studying segmental acoustic phenomena, {regardless of the number of phones in each word (i.e., the duration of a phone does not vary with the number of phones in the word, as is the case in human-produced speech \cite{dauer_1983})}. In this spirit, the $f_0$ was also kept constant, and voicing was enabled for all sonorant segments. The entire process ran on a server, systematically generating a WAV file for each word in the dictionary, coding each sample with 16 bits and producing 48,000 samples per second of audio. {For each phone sequence (word),} we synthesize two tokens, one using the adult-male configuration and one using the adult-female configuration of the Pink Trombone. The segmental articulatory targets are {identical across both generated forms,} and only the sex-specific anatomical parameters differ. This yields a parallel lexicon where every word is available for both speaker sexes.

\subsection{Algorithmic Landmark Placement}

In this context, the generation of annotations is a deterministic process that maps {a word's phonetic specification} to a precise sequence of landmark events. This procedure relies on a predefined set of rules that dictate the temporal placement of landmarks relative to the fixed duration assigned to each phoneme in a sequence. This rule-based approach ensures perfect consistency across the entire database. The same rule-based schedule is applied identically to both the male and female configurations, so that landmark labels are fully comparable across speaker sex.

The specific placement rules are defined based on the phoneme's manner of articulation:

\begin{itemize}
    \item \textbf{For Vowels and Glides:} These sonorant events are characterized by a relatively stable, open vocal tract configuration. The landmark ($<$V$>$ for vowels, $<$G$>$ for glides) is placed at the temporal midpoint of this stable articulatory posture. {Because the movement of the articulators for vowels and glides is continuous rather than abrupt, this midpoint is an approximation: the $<$V$>$ landmark targets the point of maximal vocal tract opening, and the $<$G$>$ landmark targets the point of maximal constriction.} {Thus this instant corresponds to an estimation of the point} of maximal vocal tract opening for vowels, resulting in a peak of acoustic energy, and the point of maximal constriction for glides, resulting in an energy minimum between two peaks. {Liquids (e.g., /l/, /r/) are likewise represented as glide-like constriction events ($<$G$>$), while their phone identity is preserved in the metadata.}

    \item \textbf{For Nasals:} These consonants require a specific articulatory configuration: a complete oral closure combined with an open velopharyngeal port. {These landmarks} are placed at the boundaries of this event. The Nasal Closure ($<$Nc$>$) is placed at the exact {moment that the synthesizer's} oral tract closes while the nasal passage parameter is active, and the Nasal Release ($<$Nr$>$) is placed at the moment the oral closure is released and the nasal passage inactive.

    \item \textbf{For Fricatives:} The acoustic signature of a fricative is generated by {a specific noise source} within the PT when an articulator is narrowing the vocal tract. The Fricative Closure ($<$Fc$>$) landmark is placed at the precise time sample of this noise source's activation, and the Fricative Release ($<$Fr$>$) is placed at the exact moment of its deactivation.

    \item \textbf{For Stop Consonants:} The Stop Closure ($<$Sc$>$) landmark is placed at the moment the vocal tract is completely obstructed. The Stop Release ($<$Sr$>$), representing the instantaneous burst, is placed at the moment the vocal tract {opens enough to permit} air flow.

\end{itemize}

This programmatic approach ensures that the annotations are not inferred from the acoustic output but are, by definition, {predictable from} the underlying articulatory schedule. The process yields a dataset with high-precision deterministic ground truth labels for each synthesized word, establishing an unambiguous reference for future research.

\section{Results: Database and Visualization Tool}
\label{sec:results}

\subsection{The ALLIE-PT Database}
The result of this process is the ``Articulatory Landmark Lexicon of English - Pink Trombone'' (ALLIE-PT) database. It is publicly available online\footnote{\url{https://huggingface.co/datasets/mcamara/all-words-in-english-with-pink-trombone}}.
It contains over 200,000 WAV files, representing a vast portion of the English lexicon. The audio format is 48 kHz, 16-bit PCM mono WAV. A corresponding JSON file for each audio file with sample-level landmark timestamps and articulatory labels.

Each entry is indexed by a word id (alphabetical order) and contains: (i) the corresponding audio waveform, (ii) an utterance JSON with the Pink Trombone keyframes ({phone labels}, timing in seconds, and control parameters such as tongue.index/diameter, frontConstriction.index/diameter, backConstriction.diameter, tenseness, loudness, tract length, $f_0$), and (iii) a landmarks JSON listing time-stamped acoustic events. Landmark types follow the updated versions of \cite{stevens_toward_2002} framework (e.g., Sc/Sr for stop closure/release, Fc/Fr for fricatives...). In the Hugging Face release, utterance and landmarks are stored as JSON-formatted strings alongside the audio object. Helper code is provided to parse them. More details on how to use the dataset can be found in the dataset card associated to the repository.

\subsection{Landmark Distribution and Lexical Statistics}

We compiled a summary of the overall database statistics. Table \ref{tab:summary_stats} provides an overview of the corpus scale and the relative prevalence of consonantal versus vocalic/glide events.

\begin{table}[htbp]
\caption{Overall statistics of the generated landmark database (one sex).}
\centering
\begin{tabular}{lr}
\hline
\textbf{Statistic} & \textbf{Value} \\
\hline
Total Words Processed & 115,487 \\
Total Landmarks Generated & 1,100,803 \\
Total Consonantal Landmarks & 676,646 \\
Total Vocalic/Glide Landmarks & 424,157 \\
\hline
\textbf{Consonantal-to-Vocalic Ratio} & \textbf{1.595} \\
\hline
\end{tabular}
\label{tab:summary_stats}
\end{table}

A key finding is the Consonantal-to-Vocalic {(i.e., the ratio of consonantal landmark tokens to vocalic/glide landmark tokens)} landmark ratio of approximately 1.6. This value reflects the canonical structure of our database, where most consonantal phonemes are realized as a landmark pair (e.g., a Stop Closure, $<$Sc$>$, and a subsequent Stop Release, $<$Sr$>$), while vowels and glides are marked by a single event. Therefore, while consonantal landmarks are more frequent, the underlying count of sonorant phoneme nuclei (vowels and glides) is actually more prominent in the English lexicon than consonantal phonemes.

Table \ref{tab:freq_stats} presents the absolute frequency of the eight primary landmark types. As expected, the Vowel landmark ($<$V$>$) is the most frequent event, reflecting its role as the nucleus of syllables. Events related to Stop consonants ($<$Sc$>$, $<$Sr$>$) are the most common among consonantal landmarks, indicating the prevalence of plosives in English.

\begin{table}[htbp]
\caption{Absolute frequency of each landmark type across the lexicon.}
\centering
\begin{tabular}{lr}
\hline
\textbf{Landmark Type} & \textbf{Frequency} \\
\hline
V (Vowel) & 279,980 \\
Sc (Stop Closure) & 153,181 \\
Sr (Stop Release) & 153,181 \\
G (Glide) & 144,177 \\
Fc (Fricative Closure) & 99,016 \\
Fr (Fricative Release) & 99,016 \\
Nc (Nasal Closure) & 86,126 \\
Nr (Nasal Release) & 86,126 \\
\hline
\end{tabular}
\label{tab:freq_stats}
\end{table}

\subsection{Phonotactic Patterns}

To investigate {the landmark structure of English (i.e. the phonotactic structure at the level of these acoustic cues)} from an event-based perspective, we analyzed the frequency of landmark bigrams (sequences of two adjacent landmarks). Table \ref{tab:bigram_stats} lists the most common bigrams, which reveal the dominant {articulatory/acoustic event sequences} in the language.

The most frequent bigrams are the definitional pairs for consonants, such as $<$Sc$>$-$<$Sr$>$, confirming the internal consistency of the data. {It should be noted that the $<$Sc$>$ landmark at word onset is placed by the articulatory schedule, even though in isolated words there is typically no preceding acoustic energy from which to detect an abrupt closure. This is an inherent property of single-word synthesis.} More revealing are the transitional patterns. The sequence $<$Sr$>$-$<$V$>$ is the most common non-definitional bigram, representing the canonical CV syllable onset (e.g., the transition from {[t] to [o\textipa{U}]} in ``toe''). Conversely, sequences like $<$V$>$-$<$Nc$>$ and $<$V$>$-$<$Sc$>$ represent {a Vowel-to-Consonant transition typical of} the VC syllable coda (e.g., the transition in ``an'' or ``act'').

{The high frequency of $<$V$>$-$<$G$>$ and $<$G$>$-$<$V$>$ bigrams reflects the prevalence of diphthongs and liquids in the English lexicon.} Finally, the presence of sequences like $<$Fr$>$-$<$Sc$>$ (e.g., the transition from /s/ to /t/ in ``aster'') demonstrates the model's ability to capture consonant clusters. {Table~\ref{tab:bigram_stats} is a descriptive phonotactic summary, not a claim that bigrams uniquely identify words or place of articulation: a single bigram such as $<$V$>$-$<$Sc$>$ occurs in both ``at'' and ``act'', with the distinction carried by the full landmark sequence, timing, phone labels, and articulatory keyframes.}

\begin{table}[htbp]
\caption{Top 10 most frequent landmark bigrams and their linguistic interpretation.}
\centering
\resizebox{\columnwidth}{!}{%
\begin{tabular}{cccc}
\hline
\textbf{Bigram} & \textbf{Frequency} & \textbf{Definition} & \textbf{Example} \\
\hline
$Sc$-$Sr$ & 153,181 & Definitional for any stop & ``\textbf{t}op'' \\
$Fc$-$Fr$ & 99,016 & Definitional for any fricative & ``\textbf{s}o'' \\
$Sr$-$V$  & 90,698  & CV syllable onset & ``\textbf{toe}'' \\
$Nc$-$Nr$ & 86,126  & Definitional for any nasal & ``\textbf{n}o'' \\
$V$-$G$   & 84,490  & Diphthong / Off-glide & ``b\textbf{oy}'' \\
$G$-$V$   & 80,048  & Glide onset followed by vowel & ``\textbf{ye}s'' \\
$V$-$Nc$  & 60,992  & VC coda (nasal) & ``\textbf{an}'' \\
$V$-$Sc$  & 54,446  & VC coda (stop) & ``\textbf{act}'' \\
$Fr$-$V$  & 45,338  & CV syllable onset & ``\textbf{fo}r'' \\
$V$-$Fc$  & 39,164  & VC coda (fricative) & ``\textbf{ash}'' \\
\hline
\end{tabular}%
}
\label{tab:bigram_stats}
\end{table}

\subsection{Visual Validation of the Articulatory-Acoustic Link}

\begin{figure}[tbp]
\centerline{\includegraphics[width=.8\columnwidth]{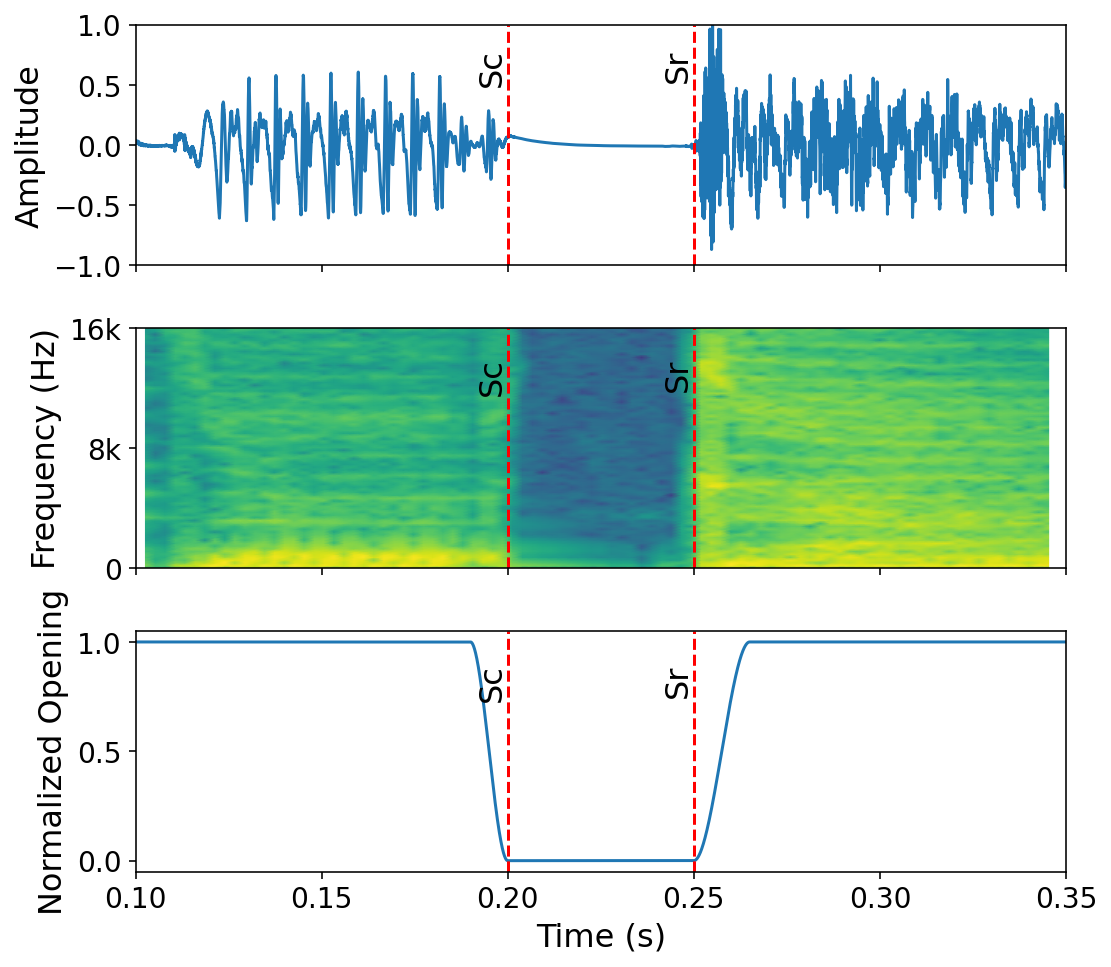}}
\caption{Visual validation of the ground-truth landmark generation for the {synthesized VCV} sequence {[ate]}. The three panels show the alignment between the articulatory command for a stop consonant (bottom), the resulting acoustic events in the spectrogram (middle), and the waveform (top). The dashed red lines indicate the precise, algorithmically-placed locations for the Stop Closure ($<$Sc$>$) and Stop Release ($<$Sr$>$) landmarks.}
\label{fig:validation_example}
\end{figure}

To demonstrate the direct correspondence between the synthesizer's internal state and the generated landmark annotations, Figure \ref{fig:validation_example} illustrates the synthesis {of the VCV sequence} [ate]. The figure presents three synchronized visualizations: the synthesized audio waveform, its corresponding log spectrogram, and the underlying ``Articulatory Opening'' parameter from the synthesizer, which serves as our ground-truth physical state. A normalized opening value of 1.0 represents an open vocal tract (as in a vowel), while 0.0 represents a complete closure.

The figure clearly shows the two primary events associated with the stop consonant /t/:

\begin{enumerate}
    \item \textbf{Stop Closure ($<$Sc$>$):} At approximately 0.20 seconds, the articulatory opening parameter drops to 0.0, representing the physical action of a complete oral closure. The landmark $<$Sc$>$ is placed at this exact instant. The direct acoustic consequence is immediately visible in the other two panels: the periodic waveform of the vowel /a/ ceases, and the formant structure in the spectrogram disappears, creating a silent interval known as a ``stop gap''.
    \item \textbf{Stop Release ($<$Sr$>$):} At approximately 0.25 seconds, the articulatory opening parameter returns to 1.0, modeling the release of the closure. The $<$Sr$>$ landmark is precisely aligned with this command. The physical result is a prominent burst transient, visible as a sharp vertical spike of broadband energy in the spectrogram, which is immediately followed by the formant structure of the vowel /e/.
\end{enumerate}

This one-to-one mapping between a programmed articulatory gesture and a perfectly aligned landmark label is the core principle of our data generation methodology. While this figure illustrates the case for stop consonants, the same relationship holds for all other landmark types detailed in Table \ref{tab:landmark_cues}. For example, a Nasal Closure landmark ($<$Nc$>$) {corresponds to the exact moment that the synthesizer's oral closure occurs while the velopharyngeal port is open}. This visual evidence confirms that our database provides a true, physically-grounded set of annotations, where each landmark is a direct record of the underlying articulatory event that caused it.

\subsection{Objective Intelligibility Evaluation}

A quantitative validation was performed to confirm the general intelligibility of the generated lexicon using the Short-Time Objective Intelligibility (STOI) metric \cite{taal2011algorithm}. STOI is a widely-used objective measure {ranging from -1 to +1} that correlates well with human speech perception by comparing a degraded or synthetic signal to a clean reference. For this evaluation, a phonetically-balanced subset of the Harvard sentences was selected \cite{rothauser1969ieee}, ensuring that all chosen words were present in our generated database.

To create the reference signals, a male native English speaker recorded these sentences. The speaker was instructed to mimic the aprosodic, flat-intonation style of the PT synthesis while maintaining correct and clear pronunciation. This approach ensures that the STOI comparison focuses primarily on phonetic accuracy and articulatory quality, rather than {prosodic or other differences}.

\begin{figure}[bp]
\centerline{\includegraphics[width=0.75\columnwidth]{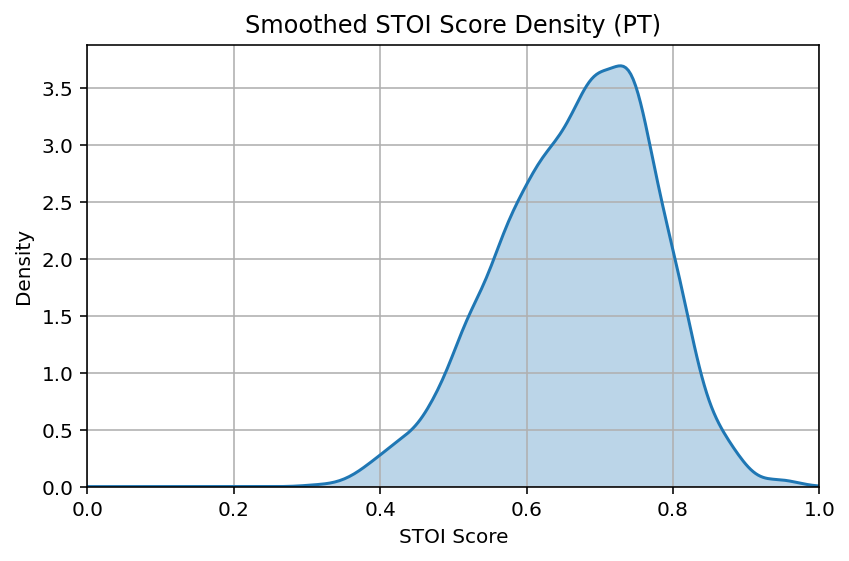}}
\caption{Smoothed density plot of STOI scores for a phonetically-balanced subset of the synthesized database, compared against clean human recordings.}
\label{fig:stoi_density}
\end{figure}

The results of this analysis are presented in Figure \ref{fig:stoi_density}. The distribution of STOI scores is unimodal, with a clear peak and average value around 0.75. According to established interpretations of the metric, a STOI score in this range represents a good (not excellent) level of intelligibility for synthetic or processed speech \cite{rothauser1969ieee}. The presence of a tail towards lower scores is acknowledged and attributed to the inherent limitations of the simplified physical model, which may struggle with certain complex phonetic sequences. Conversely, the {substantial} portion of the distribution above 0.8 demonstrates that many synthesized words are highly intelligible. This quantitative evaluation confirms that the database possesses a sufficient level of intelligibility to be a valid resource for perceptual studies and potential machine learning applications.

{In addition, by highlighting the aspects of the synthesized word forms based on landmark cues only that do not yet correspond to naturally produced human speech, this database provides a usefully transparent tool for investigation of additional cue patterns required for the synthesis of natural-sounding speech.}

\subsection{Interactive Visualization Tool}
To accompany the database, we developed a web-based interactive tool. This tool allows users to input any English word (or a phonetic transcription). It then generates the speech and visualizes the vocal tract's movement, the output waveform, the spectrogram, and the landmark locations in real-time. This serves as a pedagogical and research tool, clarifying the link between articulation and acoustics. The demo is available online\footnote{\url{https://github.com/MateoCamara/pink-trombone-demos}}.

\section{Discussion}
\label{sec:discussion}

A key aspect of this contribution is the deliberate decision not to benchmark existing landmark detectors with the generated dataset. Such systems are designed for natural speech, typically within a probabilistic framework. Our database, by contrast, is based on the fundamentals of the landmark theory, in which the ground truth is a direct record of {the acoustic landmarks and their underlying articulatory commands}. The work described here is therefore not a test set for current tools but an unambiguous training and development resource for the next generation of landmark-based models.

The choice of the PT synthesizer was central to this goal. While it is a simplification of human anatomy {(and thus the speech generated by this initial version of the ALLIE-PT method is expected to be of imperfect quality)}, its computational efficiency and transparency are features that enable the creation of {a canonical minimally prosodic lexicon}. The current database, with its {fixed phone durations} and flat $f_0$, provides a clean baseline for studying segmental phenomena, free from prosodic variability. This controlled nature, however, also highlights a clear path for future work: incorporating prosodic models to control timing and intonation would allow for the generation of more naturalistic speech, further expanding the database's utility. An additional advantage of the present corpus is that every lexical item is available for two adult speaker-sex configurations with controlled anatomical differences. This makes it possible to study how acoustic landmark patterns vary (or remain invariant) with vocal-tract length and sex-specific anatomy, while keeping the underlying articulatory schedule and lexical content strictly matched. Such parallel male–female data are difficult to obtain from natural speech without confounds such as prosody, speaking style, or recording conditions.

{A further direction for future work is the adaptation of the synthesizer to model a child vocal tract, whose shorter tract length, different formant ranges, and distinct articulatory dynamics would allow the database to support studies in speech development and pediatric clinical phonetics.}

Moreover, the methodology presented is language-agnostic and holds particular promise for under-resourced languages where collecting and manually annotating large speech corpora is time-consuming. This reinforces the role of landmarks as a unifying concept that bridges speech production, and perception, enabling controlled studies in analysis-by-synthesis. Consequently, this work opens several key research avenues. The database is ideally suited for training landmark-based ASR systems that can learn the invariant features of speech events. It also provides an ideal standard for developing and evaluating novel landmark detectors, and its extension to other languages will facilitate large-scale, quantitative comparisons of phonotactic structures across language families.

{An important caveat is that the temporal relationship between articulatory onsets/offsets and their acoustic consequences is not always instantaneous or precisely aligned. In natural speech, the acoustic manifestation of an articulatory event may lag behind or lead the physical gesture due to the biomechanical properties of the articulators and the aerodynamic conditions in the vocal tract. For example, research on nasality has shown that the acoustic onset of nasalization can lag behind the opening of the velum \cite{vaissiere_1988}. In the PT framework, landmark placement tracks the articulatory command directly (i.e., the moment the control parameter changes), which represents an idealized, deterministic mapping. While this provides an unambiguous annotation source, users of the database should be aware that in natural speech, the acoustic-to-articulatory timing relationship is more complex and variable.}

{A related limitation is that the current database assigns a single canonical articulatory target to each phone, modeling only limited aspects of context-conditioned allophonic variation. In natural speech, phones exhibit substantial variation depending on their phonetic context that will not always be captured with the phone symbols of the IPA, such as the range of acoustic cue patterns associated with the symbol for a flap \cite{yun_2025_korean_labeling}. The PT synthesis uses a single target for each phone regardless of context, which means that some cue pattern distinctions are not captured~\cite{choi_1997_asa_poster}. Incorporating context-sensitive cue pattern variation is a promising direction for future versions of the database.}

{It is also worth clarifying the sense in which we use the term ``ground truth.'' Because the articulatory commands in the PT are fully specified and deterministic, the landmark annotations are unambiguous: the system knows exactly when each articulatory gesture was commanded, and the landmark is placed at that instant. This contrasts with annotations derived from the acoustic signal, where the precise location of a landmark event may be ambiguous due to coarticulation, noise, or gradual transitions. Our ``ground truth'' thus refers to the deterministic nature of the annotation source (the articulatory schedule), not to a claim of perceptual primacy or acoustic invariance.}

\section{Conclusions}
\label{sec:conclusion}

In this work, we have presented a methodology for creating a large-scale, physically-grounded acoustic landmark database by using generative synthesis. {The quality of the synthesized speech may not be comparable to that obtained with other techniques currently used in the state of the art, but that was not the goal. Instead, this work seeks to provide a large scale reference corpus for the study and validation of the landmarks theory, using a computationally efficient and transparent framework that can facilitate a component-by-component buildup of increasingly natural sounding synthesized speech.}

Using the {physically and articulatory grounded} PT synthesizer, we produced a dataset containing synthetic speech records over a lexicon of more than 230,000 English words, each rendered for both an adult male and an adult female vocal-tract configuration. The landmark annotations were placed algorithmically to {closely} correspond to their underlying articulatory commands. A statistical analysis of this corpus revealed a consonantal-to-vocalic landmark ratio of approximately 1.6, offering a new quantitative measure of the articulatory density in the English language.

To validate the utility of the database for perceptual tasks, an objective evaluation was performed. The synthesized speech achieved an average STOI score of 0.75 when compared against human-spoken references, confirming {a sufficient initial level of intelligibility to support subsequent studies.}

The primary contribution of this research is a foundational, publicly available reference corpus, where the ground-truth annotations are deterministic, with the ambiguities of natural speech removed. This resource is intended to serve as a standard for the development and benchmarking of the next generation of automatic landmark detectors, allowing researchers to isolate model performance from data variability. Furthermore, this work provides a robust framework for quantitative linguistic analysis, {and enables new possibilities for analysis-by-synthesis research, opening the door } {to elucidating the mechanisms involved in adjusting the articulation} { in specific ways based on adjacent sounds, speaking style and rate, and location in morphemic and prosodic structure. } {In addition, the methodology} is inherently language-agnostic and can be extended to under-resourced languages, accelerating the development of more robust and interpretable event-based speech technologies.

\section{Supplementary material}

{This manuscript is supplemented with a document identifying a set of words representing canonical cases of interest for {this approach}. The acoustic waveforms and their corresponding landmarks are also included in the supplementary material to provide a clear idea about the possibilities and extent of the corpus. It can be accessed online.\footnote{\url{https://mateocamara.com/lexi-challenging-words/}}}

\section{Acknowledgments}

The authors thank Zack Qattan for foundational work mapping English phones to the Pink Trombone articulatory configurations and the PT demos. This work was supported by the Ministry of Economy and Competitiveness of Spain under grant PID2021-128469OB-I00, and by the ``Ayuda Econ\'{o}mica Para Personal Investigador Postdoctoral 2025'' of the Fundaci\'{o}n Santander. This work was also supported by a Grant of the MISTI MIT Global Experiences.

\section{Generative AI Use Disclosure}
Generative AI tools have only been used for editing/polishing (not for writing any major part). The authors reviewed all the text.

\bibliographystyle{IEEEtran}
\bibliography{strings,refs,references_zotero}

@article{he_acoustic_2018,
	title = {Acoustic landmarks contain more information about the phone string than other frames for automatic speech recognition with deep neural network acoustic model},
	volume = {143},
	issn = {0001-4966},
	doi = {10.1121/1.5039837},
	number = {6},
	journal = {The Journal of the Acoustical Society of America},
	author = {He, Di and Lim, Boon Pang and Yang, Xuesong and Hasegawa-Johnson, Mark and Chen, Deming},
	month = jun,
	year = {2018},
	pages = {3207--3219},
}

@inproceedings{shattuck-hufnagel_robustness_2007,
	title = {Robustness of {Acoustic} {Landmarks} in {Spontaneously}-{Spoken} {American} {English}},
	booktitle = {Proceedings of the 16th {International} {Congress} of {Phonetic} {Sciences}},
	author = {Shattuck-Hufnagel, S. and Veilleux, N.},
	year = {2007},
}

@article{slifka_acoustic_2006,
	title = {Acoustic {Cues}, {Landmarks}, and {Distinctive} {Features}: a {Model} of {Human} {Speech} {Processing}},
	volume = {2},
	copyright = {Copyright (c) 2016 ECTI Transactions on Computer and Information Technology (ECTI-CIT)},
	issn = {2286-9131},
	shorttitle = {Acoustic {Cues}, {Landmarks}, and {Distinctive} {Features}},
	doi = {10.37936/ecti-cit.200622.53277},
	language = {en},
	number = {2},
	journal = {ECTI Transactions on Computer and Information Technology (ECTI-CIT)},
	author = {Slifka, Janet},
	year = {2006},
	pages = {91--96},
}

@article{stevens_evidence_1981,
	title = {Evidence for the role of acoustic boundaries in the perception of speech sounds},
	volume = {69},
	issn = {0001-4966},
	doi = {10.1121/1.386489},
	number = {S1},
	journal = {The Journal of the Acoustical Society of America},
	author = {Stevens, Kenneth N.},
	month = may,
	year = {1981},
	pages = {S116},
}

@article{liu_first_2021,
	title = {A {First} {Step} toward the {Clinical} {Application} of {Landmark}-{Based} {Acoustic} {Analysis} in {Child} {Mandarin}},
	volume = {8},
	issn = {2227-9067},
	doi = {10.3390/children8020159},
	language = {eng},
	number = {2},
	journal = {Children (Basel, Switzerland)},
	author = {Liu, Chin-Ting},
	month = feb,
	year = {2021},
	pmid = {33672507},
	pmcid = {PMC7923755},
	pages = {159},
}

@article{he_selecting_2017,
	title = {Selecting frames for automatic speech recognition based on acoustic landmarks},
	volume = {141},
	issn = {0001-4966},
	doi = {10.1121/1.4987204},
	number = {5\_Supplement},
	journal = {The Journal of the Acoustical Society of America},
	author = {He, Di and Lim, Boon Pang P. and Yang, Xuesong and Hasegawa-Johnson, Mark and Chen, Deming},
	month = may,
	year = {2017},
	pages = {3468},
}

@article{liu_landmark_1996,
	title = {Landmark detection for distinctive feature‐based speech recognition},
	volume = {100},
	issn = {0001-4966},
	doi = {10.1121/1.416983},
	number = {5},
	journal = {The Journal of the Acoustical Society of America},
	author = {Liu, Sharlene A.},
	month = nov,
	year = {1996},
	pages = {3417--3430},
}

@article{huang_investigation_2022,
	title = {Investigation of {Speech} {Landmark} {Patterns} for {Depression} {Detection}},
	volume = {13},
	issn = {1949-3045},
	doi = {10.1109/TAFFC.2019.2944380},
	number = {2},
	journal = {IEEE Transactions on Affective Computing},
	author = {Huang, Zhaocheng and Epps, Julien and Joachim, Dale},
	month = apr,
	year = {2022},
	pages = {666--679},
}

@article{juneja_probabilistic_2008,
	title = {A probabilistic framework for landmark detection based on phonetic features for automatic speech recognitiona)},
	volume = {123},
	issn = {0001-4966},
	doi = {10.1121/1.2823754},
	number = {2},
	journal = {The Journal of the Acoustical Society of America},
	author = {Juneja, Amit and Espy-Wilson, Carol},
	month = feb,
	year = {2008},
	pages = {1154--1168},
}

@article{boyce_speechmark_2013,
	title = {{SpeechMark} {Acoustic} {Landmark} {Tool}: {Application} to {Voice} {Pathology}},
	volume = {2013},
	issn = {2308-457X},
	shorttitle = {{SpeechMark} {Acoustic} {Landmark} {Tool}},
	journal = {Interspeech},
	author = {Boyce, Suzanne and Speights, Marisha and Ishikawa, Keiko and MacAuslan, Joel},
	month = aug,
	year = {2013},
	pmid = {39469446},
	pmcid = {PMC11514720},
	pages = {2672--2674},
}

@misc{zhang_auto-landmark_2024,
	title = {Auto-{Landmark}: {Acoustic} {Landmark} {Dataset} and {Open}-{Source} {Toolkit} for {Landmark} {Extraction}},
	shorttitle = {Auto-{Landmark}},
	doi = {10.48550/arXiv.2409.07969},
	publisher = {arXiv},
	author = {Zhang, Xiangyu and Liu, Daijiao and Xiao, Tianyi and Xiao, Cihan and Szalay, Tuende and Shahin, Mostafa and Ahmed, Beena and Epps, Julien},
	month = sep,
	year = {2024},
	note = {arXiv:2409.07969 [eess]},
}

@book{stevens_acoustic_2000,
	title = {Acoustic {Phonetics}},
	isbn = {978-0-262-69250-2},
	language = {en},
	publisher = {MIT Press},
	author = {Stevens, Kenneth N.},
	month = jul,
	year = {2000},
}

@misc{he_improved_2018,
	title = {Improved {ASR} for {Under}-{Resourced} {Languages} {Through} {Multi}-{Task} {Learning} with {Acoustic} {Landmarks}},
	doi = {10.48550/arXiv.1805.05574},
	publisher = {arXiv},
	author = {He, Di and Lim, Boon Pang and Yang, Xuesong and Hasegawa-Johnson, Mark and Chen, Deming},
	month = may,
	year = {2018},
	note = {arXiv:1805.05574 [cs]},
}

@article{stevens_toward_2002,
	title = {Toward a model for lexical access based on acoustic landmarks and distinctive features},
	volume = {111},
	issn = {0001-4966},
	doi = {10.1121/1.1458026},
	number = {4},
	journal = {The Journal of the Acoustical Society of America},
	author = {Stevens, Kenneth N.},
	month = apr,
	year = {2002},
	pages = {1872--1891},
}

@article{vaissiere_1988,
  title   = {Prediction of Velum Movement from Phonological Specifications},
  author  = {Vaissi\`{e}re, Jacqueline},
  journal = {Phonetica},
  year    = {1988},
  volume  = {45},
  number  = {2-4},
  pages   = {122--139},
  doi     = {10.1159/000261822}
}

@misc{thapen2017pink,
  title={Pink Trombone},
  author={Thapen, Neil},
  year={2017}
}

@article{camara2025parameter,
  title={Parameter optimisation for a physical model of the vocal system},
  author={C{\'a}mara, Mateo and Blanco, Jos{\'e} Luis and Reiss, Joshua D},
  journal={EURASIP Journal on Audio, Speech, and Music Processing},
  volume={2025},
  number={1},
  pages={27},
  year={2025},
  publisher={Springer}
}

@article{maeda1990compensatory,
  title={Compensatory articulation during speech: Evidence from the analysis and synthesis of vocal-tract shapes using an articulatory model},
  author={Maeda, Shinji},
  journal={Speech production and speech modelling},
  pages={131--149},
  year={1990},
  publisher={Springer}
}

@article{birkholz2013modeling,
  title={Modeling consonant-vowel coarticulation for articulatory speech synthesis},
  author={Birkholz, Peter},
  journal={PloS one},
  volume={8},
  number={4},
  pages={e60603},
  year={2013},
  publisher={Public Library of Science San Francisco, USA}
}

@article{taal2011algorithm,
  title={An algorithm for intelligibility prediction of time--frequency weighted noisy speech},
  author={Taal, Cees H and Hendriks, Richard C and Heusdens, Richard and Jensen, Jesper},
  journal={IEEE Transactions on audio, speech, and language processing},
  volume={19},
  number={7},
  pages={2125--2136},
  year={2011},
  publisher={IEEE}
}

@article{rothauser1969ieee,
  title={{IEEE} recommended practice for speech quality measurements},
  author={Rothauser, Ernst H},
  journal={IEEE Transactions on Audio and Electroacoustics},
  volume={17},
  number={3},
  pages={225--246},
  year={1969},
  publisher={Institute of Electrical and Electronics Engineers (IEEE)}
}

@misc{cmudict,
  author       = {Carnegie Mellon University},
  title        = {The {CMU} Pronouncing Dictionary},
  howpublished = {\url{http://www.speech.cs.cmu.edu/cgi-bin/cmudict}},
  year         = {2014},
  note         = {Version 0.7b}
}

@article{camara2023optimization,
  title={Optimization techniques for a physical model of human vocalisation},
  author={C{\'a}mara, Mateo and Xu, Zhiyuan and Zong, Yisu and Blanco, Jos{\'e} Luis and Reiss, Joshua D},
  journal={arXiv preprint arXiv:2309.14761},
  year={2023}
}

@book{international1999handbook,
  title={Handbook of the International Phonetic Association: A guide to the use of the International Phonetic Alphabet},
  author={International Phonetic Association},
  year={1999},
  publisher={Cambridge University Press}
}

@article{huilgol_2019,
  title   = {A framework for labeling speech with acoustic cues to linguistic distinctive features},
  author  = {Huilgol, Shreya and Baik, Jinwoo and Shattuck-Hufnagel, Stefanie},
  journal = {The Journal of the Acoustical Society of America},
  year    = {2019},
  volume  = {146},
  number  = {2},
  pages   = {EL184--EL190},
  doi     = {10.1121/1.5121717},
  note    = {JASA Express Letters}
}

@article{dauer_1983,
  title   = {Stress-timing and syllable-timing reanalyzed},
  author  = {Dauer, Rebecca M.},
  journal = {Journal of Phonetics},
  year    = {1983},
  volume  = {11},
  number  = {1},
  pages   = {51--62},
  doi     = {10.1016/S0095-4470(19)30776-4}
}

@article{choi_1997_asa_poster,
  title   = {Labeling a speech database with landmarks and features},
  author  = {Choi, Jeung-Yoon and Chuang, Erika and Gow, David and Kwong, Katherine and Shattuck-Hufnagel, Stefanie and Stevens, Kenneth and Zhang, Yong},
  journal = {The Journal of the Acoustical Society of America},
  year    = {1997},
  volume  = {102},
  number  = {5(Supplement)},
  pages   = {3163--3163},
  note    = {ASA 134th Meeting (Abstract 4aSC6), San Diego, CA},
  url     = {https://www.auditory.org/asamtgs/asa97snd/4aSC/4aSC6.html}
}

@article{yun_2025_korean_labeling,
  title   = {Cross-linguistic adaptation of feature-cue-based speech labeling: The case of Korean (L)},
  author  = {Yun, Suyeon and Choi, Jeung-Yoon and Shattuck-Hufnagel, Stefanie},
  journal = {The Journal of the Acoustical Society of America},
  year    = {2025},
  volume  = {157},
  number  = {6},
  pages   = {4097--4101},
  doi     = {10.1121/10.0036814}
}

@article{di2021lexical,
  title={Lexical and syntactic gemination in Italian consonants—Does a geminate Italian consonant consist of a repeated or a strengthened consonant?},
  author={Di Benedetto, Maria-Gabriella and Shattuck-Hufnagel, Stefanie and De Nardis, Luca and Budoni, Sara and Arango, Javier and Chan, Ian and DeCaprio, Alec},
  journal={The Journal of the Acoustical Society of America},
  volume={149},
  number={5},
  pages={3375--3386},
  year={2021},
  publisher={AIP Publishing}
}

\end{document}